# Electrical Structure of Biological Cells and Tissues:
## impedance spectroscopy, stereology, and singular perturbation theory

Robert Eisenberg

Department of Molecular Biophysics

Rush University

Chicago IL

October 15, 2015

Impedance Spectroscopy is a technique with exquisite resolution since it resolves linear electrical properties into uncorrelated variables, as a function of frequency. This separation is robust and most useful when the system being studied is linear (If the system is significantly nonlinear, the meaning of the parameters is not clear, and they are likely to vary substantially with conditions, making models badly posed.) Impedance spectroscopy can be combined with appropriate structural knowledge (qualitative and quantitative) based on the actual anatomy of the system being studied. The combination provides biologically useful insight into the pathways for current flow in a number of cells and tissues, with more success than other methods.

Biological applications of impedance spectroscopy are often unable to take advantage of the strengths of impedance spectroscopy since so much of biology is strongly nonlinear in its essential features, and impedance spectroscopy is fundamentally a linear analysis. There is an important special case, however, present in all cells and tissues in plants and animals, the cell membrane and its capacitance, where bioelectric properties are both linear and important to cell function. Here we take advantage of the ideal properties of membrane capacitance to make impedance spectroscopy a most useful tool for determining the electrical structure of cells and tissues [18].

The capacitance of cell membranes is close to perfect, with 90 degree phase angle as close as can be measured in the entire biologically relevant frequency range of DC to (say) $10^4$ or even $10^5$ Hz [21, 23, 26-28]. Impedance spectroscopy is a useful tool for studying cell membranes then, because linear analysis is appropriate. Cell membranes define every cell in animals and plants and so impedance spectroscopy is a tool that can be used throughout biology, exploiting the linear electrical properties of this structure. Cell membranes determine a wide range of biological function and the organization of these membranes into folded and more complex structures is a general motif in biology. Indeed, a glance at the structure of a cell in any cell biology or histology text book shows that membranes inside cells are extensively folded and organized as well. Impedance spectroscopy can help provide insight into membrane structure and function throughout biology.

Cells are not simple spheres or cylinders. Rather, they include complex membrane structures that channel electrical current to the regions of interest. These structures are important in epithelia that form the kidney and gall bladder and secretory cells [6]. These structures are crucial in muscle fibers, skeletal and cardiac, where electrical signals across membranes coordinate contraction. Without coordination, skeletal muscle cannot provide useful force [61] [24]. Without coordination the heart cannot pump blood [5]. Thus, the arrangement of membranes in skeletal muscle, and some cardiac muscle, include tubules that are invaginations of the outer cell membrane [62]. The outer membrane is a cylinder (roughly speaking) that sprouts tubular invaginations that bring current flow and electrical signals into the interior of the muscle fiber. These signals, like ALL biological signals are voltages ACROSS MEMBRANES. They allow communication of the propagating action potential of the surface membrane to the depths of the muscle fiber [17].

The tubular system of muscle have an interior accessible to the solutions outside of cells. Markers diffuse quickly into the T-system as it is called. Extracellular solutions are derived from salt water and have simple linear electrical properties. Thus, the electrical properties of the T-system are roughly those of a resistor (of the extracellular solution, more or less) in series with a membrane capacitance, both actually resembling the linear electrical circuit elements used to represent them. The capacitance is of course distributed along the series resistance of the resistance. The T-system capacitance is distributed along the T-system luminal resistance.

The resistive properties of membranes in biology is hardly ever linear [29-32]. Even worse, it is the nonlinearities that are interesting. Specialized proteins called ion channels, with a hole down their middle, carry current [48, 63]. The channels open and close in response to stimuli, and so the currents through the ensemble of channels in a cell membrane performs complex biological functions and is anything but linear. Thus, studies of linear electrical properties of cells and tissues, including muscle and its T-system, choose experimental conditions in which the number of channels open remains fixed, and the resistive properties of cell membranes are linear. Fortunately, such conditions are easy to find and enforce. If perturbing potentials are chosen that make the interior of the cell more negative than its normal resting potential (of say -80 mv, inside minus outside), and not more than say 10 mv in amplitude, the resistance of membranes is constant and the membrane system (of resistor and capacitor in parallel) is a linear circuit element that can be well analyzed by impedance spectroscopy [13-15, 18, 19, 39].

The linear properties of cells of course depend on the layout of the membranes and how they are distributed along resistances [18]. Thus, to understand linear properties in physical terms one must know the anatomy of the membranes, and one must know it quantitatively. How much membrane is there? What are the dimensions of the T-tubules? How much outer membrane is there? How much membrane is in the T-system? and so on. Without quantitative anatomy, one can only measure effective phenomenological parameters of equivalent circuits. Equivalent circuits are equivalent to useful representations only in the minds of physical scientists. Biologists want to know actual circuits with parameters of definite structures and amounts of membrane. Neglect of anatomy, particularly anatomical measurements, is a sad characteristic of many impedance measurements of biological systems.

Anatomical measurements cannot be used in an electrical circuit however without a theory. To put it baldly, the units are wrong! One must have a theory to connect the structures and their amounts and sizes to the electrical measurements made by impedance spectroscopy. A theory is needed to describe current flow and the role of structure. We see then that

**Impedance spectroscopy of biological structures is a platform resting on three pillars.**
The pillars of the platform are
    (1) Impedance measurements themselves (with their technical requirements discussed below)
    (2) Anatomical observations and measurements that provide quantitative information describing the amount of membranes and their connections to each other.
    (3) A theory (field theory or circuit theory or both) linking the electrical properties of tiny pieces of membrane (differential elements, to be precise) and extra and intra cellular space with the anatomy of the system to the impedance measurements themselves.
    (4) Estimation procedures that allow the impedance measurements, anatomical observations, and theories to evaluate properties of individual membrane systems.
This paper discusses those pillars in the context of work done long ago. The paper is meant to explain and illustrate this use of impedance spectroscopy. The paper is a guide (I hope) to how the impedance spectroscopy has been successfully applied to determine the electrical structure of several biological systems of importance [18]. I apologize to the many more recent workers whom I have slighted out of ignorance and hope that they nonetheless can find this story useful.

**Anatomical measurements**. The qualitative understanding of anatomical structures underlies any quantitative measurements of those structures, and any electrical models and theories tying those structures to impedance measurements. Humans are visual animals so the unconscious visualization of structure is natural for all of us, particularly for the subset who are visually gifted. Anatomy is an ancient subject for that reason, and only rarely is our (unconscious) visualization qualitatively incorrect, and then usually when dealing with sectioned (i.e., sliced) material. When a three dimensional structure is sliced into thin sections, reconstruction into the three dimensional structure can (and has) occasionally confused spheres and cylinders because they are both circles in slices. But in general we begin with the presumption that the qualitative anatomy of the system is known through the centuries of effort of anatomists and histologists, who are anatomists on a smaller scale, and cell and structural biologists, who are anatomists on a still smaller scale reaching nowadays to the atoms that make up the structure of membranes, channels, and proteins in general.

The obvious way to measure an anatomical structure has been used 'forever'. Slice it into sections, trace out the structure, and measure the tracing. This approach is fraught with danger. As is obvious, the tissue must be fixed and embedded in a hard sliceable material before it can be sectioned and those steps must be controlled to have minimal artifact. All this is the stuff of classical histology and is more or less under control [10]. All of this can be improved or replaced by more modern three dimensional imaging techniques, if those are used to make QUANTITATIVE calibrated measurements of membrane area and so on.

What is often overlooked is that the statistical sampling methods called stereology provide far less biased estimates of crucial membrane parameters (e.g., membrane area) than tracing methods. These stereological methods are of course not the subject of this article, but their existence is so often overlooked over the years that I must point them out [8, 9]. It is easy to show in experiments on computer generated images that stereological methods are fast and accurate, whereas tracing methods are slow and badly biased. Slow because errors abound. Biased because humans and machines trace in ways that seem natural but are in fact deterministically skewed.

Stereology depends on the projection of an image onto a grid (usually rectangular) of LOW resolution with spacing between lines much larger than one might guess, and very much larger than that necessary to resolve the details of the membranes, or even their location. Resolution has to be enough only---I repeat ONLY---to allow identification of the grid point: is it inside or is it outside the structure? The loss of information in any one image is very large. BUT A LARGE NUMBER (say 500) of images of sections are needed IN ANY CASE to sample the three dimensional structure successfully. When the low resolution sampling is done of these number of images, the resulting ensemble estimates in fact have (more than) acceptable variance. And most importantly they have much less bias than tracing methods. It might be thought that making this number of observations is unacceptably tedious, but in fact the experience of many investigators show the tedium and work involved is much less than in say typical experiments of biophysics or molecular biology.

There are many modern visualization methods that allow direct measurement of tiny membrane structures, including the tubular systems of muscle, in unfixed material. These measurements are very much more reliable than measurements made in fixed and sliced material, because the fixation process introduces large, unknown, and irreproducible changes in volume

However, the stereological methods developed long ago remain very good, and probably the best method to measure membrane area. Membrane area is not disturbed by fixation; membranes and their area cannot be directly visualized by high resolution microscopy. Of course, methods that tag membranes uniformly and observe the density of the tag in high resolution microscopy would be better yet.

**Impedance Measurements.** Impedance measurements in biology are most helpful when the impedance being measured is of a system of direct biological importance. When current is applied outside cells, to a whole tissue or to a suspension of cells, the impedance measured is mostly that of the extracellular space. Properties of the extracellular space are not central to most biological function although of course they can be useful assays of malfunction (in clinical medicine) and useful adjuncts to more meaningful measurements.

Impedance measurements discussed here force the applied current to flow across membranes, through membrane capacitance and ion channels that form a conductive pathway in parallel with the membrane capacitance [11, 22, 41, 64, 67-69]. Current is forced to flow in these paths because current is applied to a tiny probe inserted into a cell, in classical experiments, or in current applied inside cells by the patch clamp method [25]. The current flows out of the probe through the cytoplasm of the cell, across the cell membrane, which is by far the largest impedance to flow, through specialized structures just outside the membrane, and then into the extracellular space and the electrode in the extracellular space, the bath electrode, which forms the other side of the current injection circuit.

With care (and some luck) probes can be inserted into most cells larger than say $10^{-5}$ m in diameter, and that includes most cardiac and skeletal muscle, although many cells do not fall into this category. The damage and perturbation produced by one probe can be assessed by inserting one probe for recording and assessment and then inserted another probe. These errors are kept well below 2% in the successful experiments discussed here.

Measurements of impedance are made in the biologically relevant frequency range of say 0.1 Hz to say 10,000 Hz where most biological function occurs. Originally, measurements were made using sinusoids and indeed much impedance spectroscopy is still done that way. Reading the engineering literature [4, 49] showed clearly, however, that measurements could be made much more quickly if the sum of sinusoids was used as the perturbing signal [39]. Since biological cells impaled with probes are dying preparations (typically lasting only an hour or so), it is important to make measurements as rapidly as possible. The rule of thumb in the engineering literature (which is easy to verify with simulations or breadboard experiments) is that the time to take a broadband measurement with the optimal sum of sinusoids is determined by the lowest frequency sinusoid. The higher frequency measurements come for free, in the sense that they do not take (significant) extra time, although of course the equipment to perform such measurements is anything but free!

This is not the place to derive or justify these statements. They have been textbook material for more than fifty years [4, 49], even if not widely known outside electrical engineering. The sum of sinusoids mentioned above is made with random phase (to keep the amplitude of the sum under control) and so the signals are stochastic requiring some knowledge of stochastic signal processing. This is knowledge that is not wide spread or trivial to master, so it is understandable that stochastic signals have not been widely used. They have great advantages nonetheless [39].

One should point out a particular artifact that is easy to make in stochastic analysis. It is tempting, even seductive to estimate a transfer function by taking a numerical estimate of a Fourier transform of the output of a system and dividing it (in the complex domain) by a numerical estimate of the Fourier transform of the input. This estimate however is so gravely flawed that is not usable, in the presence of essentially any nonlinearity or contaminating noise, as explained in great detail in the books of Otnes and Enochson [4, 14, 49]. If this estimation procedure is tried on a mock circuit, with added noise, one will find that the mean value of the estimate changes with the noise level, but the variance of the estimate does not. The mock circuit of course does not change with the level of added noise, so the qualitative properties of the estimate are wrong. If one fails to check the procedure on a mock circuit, the error can be overlooked because most workers look for variance as a sign of noise contamination, and in this case the sign of noise contamination is a bias in the mean value withOUT an increase in noise.

**Measurement Difficulties.** Probes small enough to insert into cells with acceptable damage and reproducibility are small and have high impedance even if they are micropipettes filled with nearly saturated salt (e.g., 3M KCl). At frequencies of even $10^4$ Hz, stray capacitances of even 0.1 pF are important. Thus great care must be taken in the construction of setups, analysis and treatment of errors, and calibration. This process was begun in [11, 22, 64], and extended dramatically in [41, 67-69], where attention to detail and use of a new recording amplifier allowed a calibrated bandwidth of at least $10^5$ Hz, as checked in an actual setup making biological recording. The key was the use of an inverting amplifier circuit in which the current through the micropipette is measured as a replica of the potential within the cell. The resistance of the micropipette is quite pure (because it is lumped in a tiny part of the micropipette, just a few micrometers long). This circuit has optimal noise performance as well, which was later recognized and exploited brilliantly by Neher and Sigworth [65, 66], in the electronics which Neher used in his patch clamp work.

Calibration was performed by injection of current through an AIR capacitor: a physical capacitor proved unreliable because even finger prints seriously changed the phase angle of current through it (because of the parallel conductance of finger grease). These realities proved to be the key to the later development of the integrating head state which has provided the optimal electronics for recording the picoamps of current through single channels in the AxoPatch amplifier [34-36, 56, 59] still the leading amplifier in 2015 after nearly 30 years on the market, now sold by Molecular Devices Inc.

**Future Measurements**.  The much lower impedance (by a factor of at least 5 times, without further fiddling) of patch pipettes should allow much wider bandwidth recording which might, or might not prove to lead to more biological insight. One cannot be sure until one addresses biological questions specific to a particular cell, tissues, and preparation.

**Interpreting Impedance Spectroscopy.** The results just mentioned all depend on the interpretation of the impedance spectrum. This interpretation requires a specific theory linking the measurement to the structure of the system in question, including the experimental setup and artifacts in the theory, and measurements of the anatomical parameters of the structures involved.

These requirements are often 'swept under the rug' because they require extensive work to fulfill by using an 'equivalent circuit' to replace the theory. The equivalent circuit is an essential part of the analysis to be sure, and the insight required to understand its simplifications prove to be most useful, but it is not enough to allow the full use of the wonderful resolution of impedance measurements.

The first step in extending the equivalent circuit analysis is the understanding of how the equivalent circuit approximates the electrodynamics of the biological structure and recording apparatus. The artifacts of the recording apparatus have been discussed above and enumerated and analyzed definitively (it seems to me in my prejudiced opinion) in the literature, and so will not be discussed further here [12, 68].

The relation of the equivalent circuit to the electrodynamics of the system was studied by solving Poisson's equation by several authors, but the simplification to the usual equivalent circuit only became apparent when series expansions of the solution of the Poisson equation, and the Poisson equation itself were developed[3, 12, 16, 33, 51-54].

These expansions provided wonderful insight [51]: the usual equivalent circuit (of a uniform potential for a finite cell, like a sphere, or of a transmission line for a long cylinder) arose as a separate term from the point source effects (i.e., the terms that vary with angular coordinates). Indeed, the beauty and power of the analysis (by matched asymptotic expansions for example) led applied mathematicians to use this work as a teaching example of the power of singular perturbation theory.

**Fitting Data** Once the impedance spectroscopy data is available (as plots of amplitude and phase vs frequency, for example) it is necessary to extract the parameters of the model that fits the data. The least squares fitting procedure proved robust and fast, even with ancient computers of the 1970's, and able to do the job. In fact, fitting to biological data was one of the first applications of the Brown Denis version of the Levenberg Marquardt minimization technique, now widely used and enshrined in MATLAB, thanks to the help of many coworkers.

Just as important as fitting the data, is determining the significance of the parameter estimates. The issue is not just what is the statistical variance of a parameter estimate. Here there is an additional problem because of the correlation between parameter estimates. Perhaps changing two parameters (or more) might produce a fit nearly as good as the one chosen. This issue can be addressed by fixing one parameter at a different value from its optimal, and then comparing the quality of fit, using the F-variance ratio test (which is quite robust) in (for example) the form of the R-test used widely in crystallography.

**Results.**  Measurements proved straightforward in skeletal muscle, cardiac muscle, tendon regions of skeletal muscle, and lens of the eye [6, 7, 11, 18, 22, 37, 38, 41-43, 46, 57, 64, 67-69]. In each case the measurements of impedance provided useful information concerning the tissue. In skeletal muscle, measurements provided the best estimates of the predominant (cell) membrane system that dominates electrical properties. In cardiac muscle, measurements showed definitively that classical microelectrode voltage clamp could not control the potential of the predominant membranes, that were in the tubular system separated from the extracellular space by substantial distributed resistance. In the tendon regions of skeletal muscle, electrical properties needed to interpret extensive voltage clamp analysis of the Cambridge (UK) group were measured [1]. In the lens of the eye, impedance spectroscopy changed the basis of all recording and interpretation of electrical measurements and laid the basis for Rae and Mathias' extensive later experimental work [2, 19, 40, 42-45, 50, 55, 57-60].

Many tissues are riddled with extracellular space as clefts and tubules, for example, cardiac muscle, the lens of the eye, most epithelia, and of course frog muscle. These tissues are best analyzed with a bi-domain theory [20, 42, 43] that arose from the work on electrical structure described here. There has been a great deal of work since then on the bi-domain model admirably reviewed in [47] and this represents the most important contribution to biology of the analysis of electrical structure in my view.

**Future Perspectives:**  Science is subject to fashions like all human behavior. The electrical structure of cells and tissues is out of fashion as biologists move to understand the ion channels so important for biological function. These channels function in a biological environment determined in large measure by the electrical structure of the tissues in which they are found. Computational tools are available to compute the properties of channels in the electrical structures in which they are found. Impedance spectroscopy will come back into fashion as scientists realize there are few other techniques with the resolution necessary to determine the electrical structure and how it modulates the current flows driven by ion channels.